\documentclass[preprint,showpacs,preprintnumbers,amsmath,amssymb,prb,groupedaddress]{revtex4}
\usepackage{graphicx}
\usepackage{dcolumn}
\usepackage{bm}
\usepackage{booktabs}
\usepackage{multirow}
\usepackage{color}
\begin{document}
\title{Analytic treatment of the thermoelectric properties for  two coupled  quantum dots threaded by magnetic fields }

\author{G. Menichetti$^{1}$,  G. Grosso$^{1,2}$,  and G. Pastori Parravicini$^{1,3}$}
\affiliation{$1$ Dipartimento di Fisica ``E. Fermi'', Universit\`{a} di Pisa, Largo Pontecorvo 3, I-56127 Pisa, Italy}
\affiliation{$2$ NEST, Istituto Nanoscienze-CNR, Piazza San Silvestro 12, I-56127 Pisa, Italy}
\affiliation{$3$ Dipartimento di Fisica ``A. Volta'', Universit\`{a} di Pavia, Via A. Bassi, I-27100 Pisa, Italy}

\begin{abstract}
Coupled   double quantum dots (c-2QD) connected to leads have been widely adopted as prototype model systems to verify  interference effects on quantum transport at the nanoscale.
We provide here an analytic study of the thermoelectric properties of  c-2QD systems pierced by a uniform magnetic field. Fully analytic and easy-to-use expressions are derived for all the kinetic functionals of interest.
Within the  
Green$'$s function formalism, our  results allow  a simple inexpensive procedure for the theoretical description of the thermoelectric phenomena for different  chemical potentials and temperatures of  the reservoirs, different threading magnetic fluxes, dot energies and interdot interactions; moreover they provide an intuitive guide to parametrize the system Hamiltonian for the design of best performing realistic devices. We have found that the thermopower $S$ can be  enhanced by more than ten times    and the figure of merit  $ZT$ by more than hundred times by the presence of a threading magnetic field. Most important, we show that the magnetic flux increases also the performance  of the device under maximum power output conditions.

 \end{abstract}
 \pacs{73.63.kv, 85.35.Ds, 85.35.Be}

\maketitle

\section{INTRODUCTION}

 Quantum dot systems have attracted  enormous interest as workable thermoelectric   device candidates for the study  of electronic and thermal quantum transport at the nanoscale.
The origin of such an interest both from the theoretical  and the experimental  side, resides in the potential they offer, as artificial nanoscale junctions, to explore a large variety of thermoelectric effects.
{{ Relevance of nanostructures  as performing  energy harvesting  devices was envisaged  in the pioneering paper  of Hick and Dresselhaus\cite{DRESS93}. Since then nanoscale thermoelectricity  has been addressed  by an increasing  number of theoretical and experimental works; a perspective of the field can be found in the focus point collection in Ref.[\onlinecite{SANCEZ14}].}}
In particular,   interference Ahronov-Bohm\cite{BROGI15,KANG04,KUBA02,FAZIO11}, Fano \cite{ORELLANA03,SILVA12,WIERB11,GARCIA13},  Dicke \cite{WANG13,ORELLANA94}  and {{ Mach-Zehnder~\cite{HOFER15,Samuelson17}  effects}}, inter- and intra-dot correlation effects\cite{BULKA04,SIERRA16}, coherent transport modification by external magnetic fields 
and gate voltages.\cite{BAI04,LIU10, PYL10}, have been exploited to control the performance of thermoelectric heat devices.

The system composed by two single-level quantum dots coupled to each other  (c-2QD) via metallic leads, {{ in two terminal or multiterminal setups~\cite{RSAN15}}}, and via an interdot tunneling   are most appropriate to probe how the Hamiltonian system parameters and external conditions can be varied to optimize the energy conversion efficiency and the output power of the thermoelectric device. This is a demanding task because such parameters often play conflicting roles in the optimization process. Strategies for increasing thermoelectric performances utilizing a steep slope in the transmission function   ${\mathcal T}$(E), or its specific shape, or its resonances, have been well described in Ref.[\onlinecite{Lambert16}] where also a comparison between the thermoelectric efficiency of inorganic and organic materials is discussed.

Enhancing  thermoelectric performance   in linear regimes, requires maximization of the dimensionless thermoelectric figure of merit $ZT= \sigma S^2T/\kappa$ where $\sigma$   is the electrical conductance, $S$ the thermopower (Seebeck) coefficient, $T$ is the temperature and $\kappa=\kappa_e+\kappa_p$ is the thermal conductance (which includes electronic and lattice contributions).  
 In the search of  optimal thermoelectric response  of the device, most important quantities are its 
maximum efficiency as  thermoelectric generator,  and the efficiency at the maximum of the output power.

A crucial aspect for the evaluation of the thermoelectric response of a device, is the wide parameters range to be explored  simultaneously to determine  its optimal functioning. In this context,  the possibility of using analytic expressions for all the involved thermoelectric functions  greatly simplifies the task.   
In the literature, the analytic treatment of the c-2QD is  confined at sufficiently small temperatures by means of the Sommerfeld expansion, extended when necessary to fourth order in $k_BT$ in the evaluation of kinetic parameters.\cite{SILVA12}
{{ In the case of Lorentzian shape  of the transmission function, analytic expressions of the thermoelectric transport  coefficients have been obtained in terms  of digamma functions~\cite{SANCEZ15}. In the more complicated transmission function of  coupled double dot, we provide, in terms of trigamma functions,}}
  analytic expressions for  the relevant quantities describing the thermoelectric behavior  of a c-2QD. 
The description of the c-2QD electronic transport is performed within the Green$'$s function  framework. 
The pole structure  of the transmission function ${\mathcal T}$(E) is discussed, and the analytic expressions of the kinetic parameters, produced by ${\mathcal T}$(E), are obtained in terms  of the Bernoulli numbers and of the trigamma function~\cite{GRAD,ABRA} routinely contained in common software libraries.
We have exploited such expressions to study the variation of  Seebeck coefficient, figure of merit,  energy conversion efficiency and output power, as function of temperatures  and chemical potentials of the reservoirs, and of the magnetic field threading the  c-2QD.    In particular we focus on the thermoelectric  efficiency of the  c-2QD device,  in contact with left and right reservoirs, when it operates  at maximum  output power conditions.

We adopt the convention that  the left reservoir is the hotter one ($T_L>T_R$) while no a priori assumption is done  on the relative position of the chemical potentials $\mu_L$ and $\mu_R$ of the left and right reservoirs. 
{{ We  consider  a two-terminal quantum dot setup}}, stationary transport conditions, absence  of lattice contributions to thermal conductivity ($k\approx k_e $), and no electronic correlation effects.
The general expression for thermoelectric  transport charge current $I$ through the c-2QD, in stationary conditions,  is given by~\cite{BUT90}
 \begin{equation}
     I= \frac{- e}{h} \int_{-\infty}^{+\infty} dE
                 \, {\mathcal T} (E)  \left[ f_{L}(E) - f_{R}(E) \right]  
 \end{equation}
 where $f_{L,R}$ denote the Fermi functions of the two reservoirs.
The  electric power output (${\mathcal P}$(E) $>$ 0) is given by
 \begin{equation}
{\mathcal P} =  -I\Delta V= \frac{1}{h} \, (\mu_R -\mu_L)  \int  dE
                 \, {\mathcal T} (E)  \left[ f_{L}(E) - f_{R}(E) \right] \ .
 \end{equation}
 where {{ $\Delta V= (\mu_L-\mu_R)/(-e)$ }} is the voltage drop and $e=|e|$ is absolute value of the electron charge.

The thermoelectric efficiency of the device is given by  the ratio between the work done and the heat extracted from the high temperature reservoir:
\begin{equation}
\eta=\dfrac{W}{Q_L}=\dfrac{(Q_L-Q_R)}{Q_L} \ .
\end{equation} 

In steady state conditions the heats per unit time are the thermal currents and $W$ per unit time  is the output power ${\mathcal P}$. Then 
\begin{equation}
 \eta =  (\mu_R - \mu_L) \frac{ \int dE
                 \, {\mathcal T} (E)  \left[ f_{L}(E) - f_{R}(E) \right]   }
                 { \int dE (E- \mu_L)
                 \, {\mathcal T} (E)  \left[ f_{L}(E) - f_{R}(E) \right]  }  \ .
\end{equation} 

Expressions from (1) to (4)   depend on the thermodynamic parameters $\mu_L, T_L, \mu_R, T_R$ and by the c-2QD transmission function ${\mathcal T}$(E), and hold in the linear and nonlinear regimes.
In this paper we are interested in the linear response of the system  so that $\Delta\mu=\mu_L-\mu_R$  and  $\Delta T=T_L-T_R$ are infinitesimal quantities.
To first order in $\Delta T$ and $\Delta \mu$,  we can write 
\begin{equation*}
f_{L}(E) - f_{R}(E) = ( -  \dfrac{ \partial f_L}{\partial  E} )
                   \left[  \Delta \mu  \ + \  (E-\mu_L) \, \dfrac{ \Delta T}{T_L} \right]  \ .
\end{equation*}
Expressions (1), (2) and (4) become~\cite{NOI}
\begin{subequations}
        \begin{eqnarray}
             I_e &=& \frac{-e}{h} \int dE
                      \, {\mathcal T} (E) \, ( -  \frac{ \partial f}{ \partial  E} )
                   \left[ -e \,\Delta V + (E-\mu) \dfrac{ \Delta T}{T} \right]  
                                            \\ [3mm] 
                 {\mathcal P} &=&   \frac{1}{h} \, e \, \Delta V \int dE
                      \, {\mathcal T} (E) \, ( -  \frac{ \partial f}{ \partial  E} )
                        \left[ -e \,\Delta V + (E-\mu) \frac{ \Delta T}{T} \right] 
                                            \\[2mm] 
                  \eta &=&   e \, \Delta V  \frac{ \int dE
                      \, {\mathcal T} (E) \, ( -  \dfrac{ \partial f}{ \partial  E} )
                        \left[ -e \,\Delta V + (E-\mu) \dfrac{ \Delta T}{T} \right] }
                        {  \int dE \,   (E- \mu)  
                \,  {\mathcal T}(E) \, ( -  \dfrac{ \partial f}{ \partial  E} )
                    \left[ -e \,\Delta V + (E-\mu) \dfrac{ \Delta T}{T} \right] }  \,.
            \end{eqnarray}  
           \end{subequations}
 For convenience, in Eqs.(5) the thermodynamic parameters     $\mu_L$,   $T_L$ and the Fermi function $f_L$ are denoted  dropping the now inessential subscript $L$.

In Section II  we report   details on the c-2QD system  and its description in terms of localized functions. In Section III we provide our novel analytic   expressions  of the transport
parameters relevant  to control and design of the thermoelectric response of the c-2QD, in the linear response regime. Application of the above expressions and discussion of the results  are reported in Section IV where contour plots are reported to better evidence the energy and magnetic field  values  eventually responsible of  efficiency at the maximum output power.
 We have found that the thermopower $S$ may be enhanced by more than ten times  and the figure of merit  $ZT$ by more than hundred times due to  a threading magnetic field.   We { {red} look for} chemical potential and magnetic flux values which give  the maximum  output power and demonstrate that  the magnetic flux also increases the  corresponding  efficiency.
Section V contains our conclusions.



     \section{System description and model} 
     
          In this section we establish a localized basis model for the c-2QD 
      electronic system in contact with the left and right reservoirs, in the 
      presence of a threading magnetic field. To keep the model at the essential, 
      we make some  simplifications that could be dropped or better analyzed, 
      when necessary.
      
       Consider  a double dot electronic system, with a single orbital 
  per dot, described within the one-electron approximation in 
  the tight-binding framework.  The one-electron Hamiltonian can be 
  partitioned  in the  left lead, central device, right lead, and coupling interaction 
        \begin{equation}
                   H = H^{(left)} + H^{(dots)} + H^{(right)} + W^{(dots-leads)} \ .
        \end{equation} 
 The electronic system is schematically pictured in Fig.1, where the presence of a uniform magnetic field is also considered.
   
    \begin{figure}[h]
\begin{center}
\includegraphics{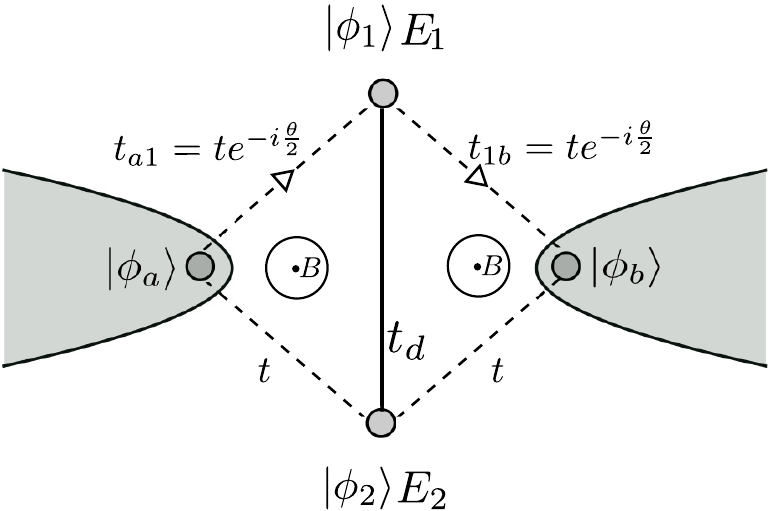}
\end{center}
\caption{Schematic representation 
   of the  double dot electronic system in a symmetric 
   environment  for the analysis of  thermoelectric properties.  
   In the absence of   magnetic field, the four hopping parameters 
   of the ring are equal to $t$  (taken as real). The magnetic field, in the chosen gauge,   
   modifies $t_{a1}\rightarrow t e^{-i\theta/2} = t_{1a}^\ast$
    and $t_{1b}\rightarrow t e^{-i\theta/2} = t_{b1}^\ast$, 
    where $\theta = 2\pi\Phi(B)/\Phi_0$,   $\Phi(B)$  is the flux of the 
    magnetic field through {{ the entire two-loop ($\phi_{a}, \phi_{1} , \phi_{b},  \phi_{2}$) plaquette}}, and $\Phi_0=hc/e$ is 
    the quantum of flux. In the case of degeneracy  $E_1=E_2=E_d$.} 
    \end{figure}

 The central device, a double dot  molecule, is described by the  
  Hamiltonian of the type in the bra-ket notations
          \begin{equation}
	    H^{(dots)} =  E_d |\phi_{1}\rangle \langle \phi_{1}| 
	                        + E_d |\phi_{2}\rangle \langle \phi_{2}|
		             + t_{d}|\phi_{1}\rangle \langle \phi_{2}| 
		                  + t_{d} |\phi_{2}\rangle \langle \phi_{1}| \ ,
         \end{equation}  
   where $E_d$ is the energy of both dots {{ orbitals $\phi_{1}, \phi_{2}$}}, and $t_d$ (supposed  real and negative) is the off-diagonal  coupling between the two dots. 
  
   For what concerns the description of  two electrodes not yet coupled to the dots, we can proceed as follows. Consider, for instance, the left lead and  
   specifically  the ``left seed state" $|\phi_a>$ that carries the coupling with the
    central  device. The effect of all the other (infinite) degrees of freedom 
   of the left electrode  are  embodied in the Green's function $g_{aa}$ on the  
   end seed state. In principle, the Lanczos procedure can be applied to 
   generate the  Lanczos chain and, then,  to determine  the Green's function       
   [see for instance Ref.[\onlinecite{SSP}]].  The same considerations  
   apply  for the  right lead.  We have
          \begin{equation}
             g_{aa}^R(E) = \langle  \phi_a| \frac{1}{E-H^{(left)} +i \eta} 
                   |  \phi_a \rangle \quad; \quad
             g_{bb}^R(E) = \langle  \phi_b| \frac{1}{E-H^{(right)} +i \eta} 
                   |  \phi_b \rangle  \ .
           \end{equation}  
    Following the   
  routinely adopted ``wide-band approximation" we  consider explicitly 
  only the imaginary part of the above Green's functions  and disregard the   
  energy dependence.  The leads are replaced by the corresponding 
  end states, with the  retarded and advanced Green's functions purely 
  imaginary quantities,  independent from energy.  In a symmetric geometrical
  environment, we have
         \begin{equation}
           g_{aa}^R(E) =  g_{bb}^R(E) \approx - i  \pi \rho   \quad , \quad
           g_{aa}^A(E) =  g_{bb}^A(E) \approx + i \pi \rho 
                                    \qquad (\rho >0) \ ,
        \end{equation}  
    where $\rho = -(1/\pi)\, {\rm Im}\, g^R$ represents the local density-of-states,
    assumed to be constant in the typical energy region 
    of actual interest.

   The coupling between leads and central device in the absence of 
   magnetic field is  represented by a loop with nearest 
   neighbor  interaction $t$ (taken as real  for simplicity). In the 
   presence of magnetic field, appropriate Peierls
    phases  are introduced. The Berry phases corresponding to the magnetic 
    field are set  on the hopping
     parameters connecting the upper quantum dot $\phi_1$ with
     the end orbitals $\phi_a, \phi_b$ of the electrodes:   
           \begin{eqnarray}
		W^{(dots-leads)} &=&   t e^{-i\theta/2)} |\phi_{a}\rangle \langle \phi_{1}| 
		                           + t e^{i\theta/2)} |\phi_{1}\rangle \langle \phi_{a}| 
		                     + t |\phi_{a}\rangle \langle \phi_{2}| 
		               + t |\phi_{2}\rangle \langle \phi_{a}|
					    \nonumber \\[2mm]
		       & + &  t e^{i\theta/2)} |\phi_{b}\rangle \langle \phi_{1}| 
		                          \, + t e^{- i\theta/2)} |\phi_{1}\rangle \langle \phi_{b}| 
		                   \,  + t |\phi_{b}\rangle \langle \phi_{2}| 
		              \, + t |\phi_{2}\rangle \langle \phi_{b}| \ .
         \end{eqnarray} 
    We have now   all the ingredients for the calculation of the Green's function
    and of the transmission function of the electronic device.  
    
\vspace{0.5cm}
\noindent {\bf A. Green's function of the degenerate double dot in magnetic fields}	
 \vspace{0.5cm}    
 
   The central part of the device is constituted by the two orbitals of the two
   quantum dots, coupled one to the other. We can use the renormalization-decimation procedure to
   fully eliminate the degrees of freedom of the leads, now represented by 
   the end seed states $|\phi_a>$ and  $|\phi_b>$ {{ [see for instance Ref.[\onlinecite{SSP}]]}}. The retarded self-energies 
   produced by the left lead on the  central device become
        \begin{equation}
                \begin{array}{l}
	    \Sigma_{11}^{R(left)}   = t_{1a} g_{aa}^{R} t_{a1} =  -i\pi \rho \,t^2
	                          	                  \\[1mm]
	       \Sigma_{12}^{R(left)}    
	          = t_{1a} g_{aa}^{R} t_{a2}  =  -i\pi \rho \, t^2 e^{i\theta/2}
	                                     \\[1mm]
	               \Sigma_{21}^{R(left)}   
	             =  t_{2a} g_{aa}^{R} t_{a1} = -i\pi \rho \, t^2 e^{- i\theta/2}
	                                            \\[1mm]
	                 \Sigma_{22}^{R(left)}  
	        =  t_{2a} g_{aa}^{R} t_{a2}  = -i \pi \rho \, t^2 \ .
	            \end{array}
           \end{equation}  
   Similar procedures can be followed for the right lead and for the advanced 
   self-energies.

     It is convenient to define  the real and positive  
     quantity  $\gamma/2 =   \pi \rho \, t^2  >0$  , that  encompasses two 
     parameters of the structure into a single one.   Using Eqs.(11),  
    the retarded (advanced) self-energy  matrix produced by the left lead 
    in the central device can   be cast in the form
              \begin{subequations}
               \begin{equation}
                    \Sigma^{R(left)}  = -i \,\frac{\gamma}{2} \left[  \begin{array}{cc}
	                  1      &     e^{i \theta/2}
	                                       \\[2mm]
	                 e^{-i \theta/2}    &   1 
                    \end{array} \right]   \quad; \quad  
                  \Sigma^{A(left)}  = i \,\frac{\gamma}{2} \left[  \begin{array}{cc}
	                  1      &     e^{i \theta/2}
	                                       \\[2mm]
	                 e^{-i \theta/2}    &   1 
                    \end{array} \right]    .        
            \end{equation} 
     (with $ \gamma >0$). Similarly, for the retarded and advanced self-energies    
     produced  by the right lead, we have
                \begin{equation}
                    \Sigma^{R(right)}  = -i \,\frac{\gamma}{2} \left[  \begin{array}{cc}
	                  1      &     e^{-i \theta/2}
	                                       \\[2mm]
	                 e^{i \theta/2}    &   1 
                    \end{array} \right]   \quad; \quad  
                  \Sigma^{A(right)}  = i \,\frac{\gamma}{2} \left[  \begin{array}{cc}
	                  1      &     e^{-i \theta/2}
	                                       \\[2mm]
	                 e^{i \theta/2}    &   1 
                    \end{array} \right]  .        
            \end{equation} 
      The total self-energies of the left and right leads  are then
           \begin{equation}
                   \Sigma^{R}  \!=\! \Sigma^{R(left)} + \Sigma^{R(right)}
                 \!=\!  - i  \gamma \left[   \begin{array}{cc}
                       1  & \!\!\! \cos (\theta/2)
                                      \\[1mm]
                     \cos (\theta/2)  &  1
                    \end{array} \right]   \  \ , \ \
                       \Sigma^{A}  
                 \!=\!  i  \gamma \left[   \begin{array}{cc}
                       1  & \!\!\! \cos (\theta/2)                                      \\[1mm]
                     \cos (\theta/2)  &  1
                    \end{array} \right] .   
            \end{equation} 
         Finally  the coupling parameters  are given by the expressions              
             \begin{equation}
                 \Gamma^{(left)}  \!=\!  i [ \Sigma^{R(left)} \!-\! \Sigma^{A(left)}] 
                      = \gamma \left[  \begin{array}{cc}
	                  1      &  \!\!\!   e^{i \theta/2}
	                                       \\[2mm]
	                 e^{-i \theta/2}    &   1 
                    \end{array} \right]         
                                \  , \
                 \Gamma^{(right)}   = \gamma \left[  \begin{array}{cc}
	                  1      &  \!\!\!   e^{-i \theta/2}
	                                       \\[2mm]
	                 e^{i \theta/2}    &   1 
                    \end{array} \right]  .       
             \end{equation}    
              \end{subequations}
    It should be noticed that the self-energies $\Sigma$ and the broadening
   parameters $\Gamma$ depend on the applied magnetic field, but are 
    completely independent  from the energy variable.  This  nice feature is a
   consequence of the wide band approximation and fosters the possibility 
   of a fully analytic treatment of transport parameters, which is a key aspect 
   of this article. 
            
      The retarded effective Hamiltonian for the double-dot in the central device,
     after the full decimation procedure of the leads, is given by
     the expression
	          \begin{equation*}
	          H^{R(eff)} =  H^{(dots)} + \Sigma^R = \left[  \begin{array}{cc}
	                  E_d      &     t_d
	                                       \\[2mm]
	                t_d    &   E_d 
                    \end{array} \right]
                           - i \gamma  \left[   \begin{array}{cc}
                       1 \ & \ \cos (\theta/2)
                                      \\[1mm]
                     \cos (\theta/2) \ & \ 1
                    \end{array} \right]  .
        \end{equation*} 
      It follows
             \begin{equation}
                  E -  H^{R(eff)}     =  \left[  \begin{array}{cc}
	                E - E_d +  i\gamma \    & \  -t_d + i \gamma \cos(\theta/2) 
	                                       \\[2mm]
	             -t_d + i \gamma \cos(\theta/2) \    & \  E - E_d +  i\gamma
                 \end{array} \right] .
             \end{equation} 
     The  inversion of the above matrix  provides the 
    retarded Green's function, represented by the symmetric matrix
	\begin{subequations}
	\begin{equation}
		G^{R}(E) = \frac{1}{E-   H^{R(eff)}} = \frac{1}  {D^{R}(E)}
              \left[  \begin{array}{cc}
	               E - E_d +  i \gamma \      & \ t_d - i\gamma \cos(\theta/2) 
	                                       \\[2mm]
	             t_d - i\gamma \cos(\theta/2) \    & \  E -E_d +  i\gamma
                 \end{array} \right]  ,
	\end{equation} 
     where
	\begin{equation}
	   D^{R}(E) =   (E  - E_d +  i \gamma )^2  
	                          -  [\, t_d - i\gamma \cos (\theta/2) \, ]^2 \ .
	\end{equation} 
	\end{subequations}
   The advanced Green's function is the hermitian conjugate of the 
   retarded one. Since the matrix $G^R(E)$ in Eq.(14) is symmetric, it follows 
             \begin{equation}
                      G^A(E) = [ G^R(E)]^\ast  \ .
            \end{equation} 
   In the present case,  the  advanced  
   Green's function is the complex conjugate of the retarded one.

\newpage
\noindent {\bf B. Transmission function of the symmetric double dot 
                                          in magnetic fields}	
 \vspace{0.5cm}    
 
     We can now proceed to the explicit calculation of the transmission 
    function  ${\mathcal T}(E)$ of the  double dots, coupled one to the other and
    immersed in  magnetic fields. Using the general Keldysh nonequilibrium 
   formalism (applicable to interacting or noninteracting systems) 
   or the Landauer-B\"{u}ttiker  procedure (specific for the latter case) 
    [see for instance Refs.\onlinecite{GOO,DATTA}], 
   we have that the transmission coefficient of the non-interacting 
   nanostructure is  given by the familiar relation
           \begin{equation}
               {\mathcal T}(E) = {\rm Tr} \left[ \Gamma^{(left)} G^R(E)
                                          \Gamma^{(right)} G^A(E) \right]  ,
           \end{equation}  
   where we have taken notice that, in the wide band approximation, 
   the left and right coupling are independent from energy.  
   
   To perform the product of the four matrices in Eq.(16), we begin to consider   
   the product of the first two matrices. Using Eq.(12d) and Eq.(14) 
   one obtains
        \begin{eqnarray}             
            & & \hspace{-0.6cm} \Gamma^{(left)} G^R(E)
                   = \frac{\gamma}  {D^{R}(E)}   \left[  \begin{array}{cc} 
	                  1      &     e^{i \theta/2}
	                                       \\[2mm]
	                 e^{-i \theta/2}    &   1 
                    \end{array} \right] 
              \left[  \begin{array}{cc}
	               E - E_d +  i \gamma  
	                        & t_d \!-\! i\frac{\gamma}{2} ( e^{i\theta/2} \!+\! e^{-i\theta/2} )
	                                    \nonumber   \\[2mm]
	                   t_d \!-\! i\frac{\gamma}{2}( e^{i\theta/2} \!+\! e^{-i\theta/2} )  
	                               & \  E -E_d +  i\gamma
                 \end{array} \right]
                                        \\[4mm]  
               & &   \!\!\!\!\!\!\! = \frac{\gamma}  {D^{R}(E)}   \left[  \begin{array}{cc}
	                      \!\!\!\!\!  E \!-\! E_d   \!-\!  i \frac{\gamma}{2}( e^{i \theta} \!-\!1 ) 
	                          \!+\! t_d e^{i \theta/2}  
	             & \!\!\!\!\!\!  e^{i \theta/2}(E \!-\! E_d) \!-\!  i \frac{\gamma}{2} 
	                           (-e^{i \theta/2}   \!+\!  e^{-i \theta/2}) \!+\! t_d
	                                 \nonumber     \\[2mm]
                      e^{-i \theta/2}(E \!-\! E_d) \!-\!  i \frac{\gamma}{2}
                                    ( e^{i \theta/2}  \!-\!  e^{-i \theta/2})  \!+\! t_d     
	                     & E \!-\! E_d  \!-\!  i \frac{\gamma}{2}(-1 \!+\!  e^{-i \theta}) 
	                             \!+\! t_d e^{-i \theta/2}
                 \end{array}  \!\!  \right]  . 
                        \end{eqnarray}
            \begin{equation}   \   \end{equation} 
         From Eq.(12d) and Eq.(15), we also have 
         \begin{equation*}    
            \Gamma^{(right)} G^A(E) = \left[ \Gamma^{(left)} G^R(E) \right]^\ast \ .
          \end{equation*}
    Multiplication of the matrix  of Eq.(17) by its complex conjugate matrix, 
    followed by the trace operation, gives  the transmission function.  
        
    After  somewhat lengthy but straight manipulations one obtains 
    the expression of the   transmission 
    function of a coupled double quantum dot   
    in a uniform magnetic field and symmetrical geometry:
        \begin{subequations}
           \begin{equation} 
              {\mathcal T}(E) = \frac{4 \gamma^2}{D^R(E)D^A(E)}
                      \left[  \cos (\theta/2) \cdot (E-E_d)  +  t_d \ \right]^2  
            \end{equation} 
    where
           \begin{eqnarray} 
                   D^{R}(E) &=&   (E  - E_d +  i \gamma )^2   
	              -  [\, t_d - i\gamma \cos (\theta/2) \, ]^2   \equiv [ D^{A}(E)]^\ast
                                           \nonumber    \\[2mm]    
                       &=&  [E  - E_d - t_d  +  i \gamma  + i\gamma \cos (\theta/2)]  
	                       [E  - E_d + t_d  +  i \gamma  - i\gamma \cos (\theta/2)]                          
                                             \qquad     \nonumber        \\[2mm] 
                       &=&  [E  - E_d  -   t_d + 2 i \gamma  \cos^2 (\theta/4)]  
	                       [E  - E_d  +   t_d  + 2  i \gamma  \sin^2 (\theta/4)] \, .                         
           \end{eqnarray}  
           \end{subequations}                             
   Whenever necessary,  some of the approximations  done for sake of simplicity 
   and for making transparent the main guidelines can be overcome at the
  modest cost  of some further manipulation. For instance the same 
  procedure can be  exploited in the case the dot levels are non degenerate, 
  or the geometric environment is non-symmetric, the magnetic field is   
  nonuniform, for multilevel dots, and other similar  situations.
  
     For instance, in the case of a non-degenerate  
   double quantu.epsm dot, with levels $E_1\neq E_2$ in a symmetric geometrical
    environment  the transmission function becomes
          \begin{subequations}
           \begin{eqnarray} 
              {\mathcal T}(E) &=& \frac{\gamma^2}{D^R(E) D^A(E)}
                         \left[ (E - E_1)^2 + (E - E_2)^2 
                              +  2\cos \theta \cdot (E - E_1)(E - E_2) \right.
                                  \nonumber \\
                  & &  \hspace{3cm}   \left.  + 4 t_d \cos (\theta/2) 
                           \cdot (2E -E_1 - E_2)  + 4 t_d^2 \, \right]  ,
              \end{eqnarray}    
      where
           \begin{equation} 
                   D^{R}(E) =   (E  - E_1 +  i \gamma )(E  - E_2 +  i \gamma )  
	                          -  [\, t_d - i\gamma \cos(\theta/2) \, ]^2  
	                          \equiv  [ D^{A}(E)]^*  \ .      
           \end{equation}   
           \end{subequations}
     In the case of degeneracy $E_1=E_2 = E_d$, one recovers back Eqs.(18).

   \vspace{0.5cm}
\noindent {\bf C. Magnetic field effects on the transmission function}	
 \vspace{0.5cm}      
    
        In the following we keep on focusing  on the degenerate double dots. 
    The deep interference effects of the  magnetic field on the transmission
  function, with  the introduction of  sharp resonances and  anti-resonances, 
  make these and similar nano-structures appealing candidates for thermoelectric   
  applications.

   According to Eqs.(18) the   transmission function of  the double quantum dot system, 
   as a function of the energy variable and  of the magnetic phase variable, 
    takes the  form
           \begin{equation} 
              {\mathcal T}(E,\theta) =  4 \gamma^2 \, \frac{
                      \left[ (E-E_d) \cos (\theta/2)   +  t_d \ \right]^2 }
              { \left[(E  \!-\! E_d  \!-\!   t_d)^2 
	                      + 4 \gamma^2  \cos^4 (\theta/4)\right]  
	              \left[(E \!-\!E_d \!+\! t_d)^2 +4 \gamma^2  \sin^4 (\theta/4)\right] } . 
              \end{equation}   
     The transmission function versus $\theta$ is periodic  with period $4\pi$, 
     corresponding to two additional flux quanta, {{ or equivalently to one flux quantum for each of the two loops of Fig.1}}.

    In the absence of magnetic fields (or  in the presence of an even 
    number of flux quanta), from Eq.(20) one obtains
                \begin{equation}
                {\mathcal T}(E,0) =   \frac{ 4 \gamma^2} 
                  { (E - E_d   -   t_d)^2  + 4 \gamma^2  } \ , 
               \end{equation}
    which is just a Lorentzian function  centered  at $E=E_d +t_d = E_d -|t_d|$, 
    the  {\it bonding} state, and effective width $\Gamma_{eff} = 2\gamma$.  
    In the presence of one  flux quantum  (or any odd integer number 
    of flux quanta) Eq.(20) gives
            \begin{equation}
                {\mathcal T}(E,2\pi) =   \frac{ 4 \gamma^2} 
                  { (E - E_d   +   t_d)^2  + 4 \gamma^2  } \ , 
               \end{equation}
     which is  a Lorentzian function  centered  at $E= E_d - t_d = E_d + |t_d|$, 
     the  {\it anti-bonding} state, and effective width $\Gamma_{eff} = 2\gamma$.
    At semi-integer flux quanta $\theta = \pi$ (or any odd integer number of $\pi$) 
    the transmission function versus $E$ takes the symmetric structure with respect to the dot energy $E_d$, with expression
          \begin{equation}
                {\mathcal T}(E,\pi) =   \frac{ 4 \gamma^2 t_d^2} 
          {[ (E  - E_d - t_d)^2  +  \gamma^2 ] [(E -E_d + t_d)^2  +  \gamma^2]  } \ . 
         \end{equation} 
      For $\gamma<< |t_d|$ (including also $\gamma \le |t_d|$) the transmission   
      function of Eq.(23) exhibits two peaks at  $\pm (t_d^2 - \gamma^2)^{1/2}$,
       and a  valley around $E=0$. The two peaks are well separated 
       if  $|t_d| >> \gamma$.

     It is of much importance to notice that, apart the special 
     values $\theta= 0, \pi, 2\pi, 3\pi$ 
     (modulus $4\pi$) discussed above, for finite values of $E$, the 
     transmission function of Eq.(20) has  a   unique zero; namely: 
           \begin{equation}
                  {\mathcal T}(E,\theta) \equiv 0  \qquad \Longrightarrow \qquad
                             E\equiv  E_d - \frac{ t_d}{\cos(\theta/2)} 
                                = E_d +  \frac{|t_d|}{\cos(\theta/2)}\ .
             \end{equation} 
   Thus the antiresonance is at the right of the anti-bonding state  for  $0<\theta < \pi$,
  while  it is at the left of the bonding  state for $\pi<\theta < 2\pi$.

   From the above discussion, {{ it is seen  how the application of the magnetic field may transform}}  a trivial unstructured    Lorentzian 
         function into a peaked-valley-peaked-valley (with zero minimum) 
         sharply structured function, 
         with much benefit in  the entailed thermoelectric properties. In general, 
         the transmission function can be qualitatively described as  the sum 
         of a Lorentzian-like curve around the bonding level and a Fano-like 
         curve around   the anti-bonding level (or vice versa, depending 
         on the applied magnetic field), with separation connected to the  coupling energy $|t_d|$.
         
     The features so far described are clearly apparent in Fig.2a, where the 
     transmission function ${\mathcal T}(E,\theta)$ is reported for various 
     values of $\theta$. For $\theta=0$ one has a simple Lorentzian 
     function centered around the bonding state. For $\theta=\pi/2$ the 
     curve shrinks around the bonding level and enlarges around 
     the anti-bonding level. For $\theta=\pi$ the curve is symmetric around 
     the dot level $|E_d|$. For $\theta=3\pi/2$ the curve appears to shift and increase
     around the anti-bonding level, and finally at $\theta=2\pi$ 
     the Lorentzian shape is recovered, now centered around 
     the anti-bonding level. In Fig.2b the full contour plot of  the 
     transmission function ${\mathcal T}(E,\theta)$ is reported.
     The information contained in Fig.2 shows clearly the 
     energy regions where the  transmission function varies rapidly so to enhance the thermoelectric  performance.\cite{Lambert16}
   \begin{figure}[h]
\begin{center}
\includegraphics{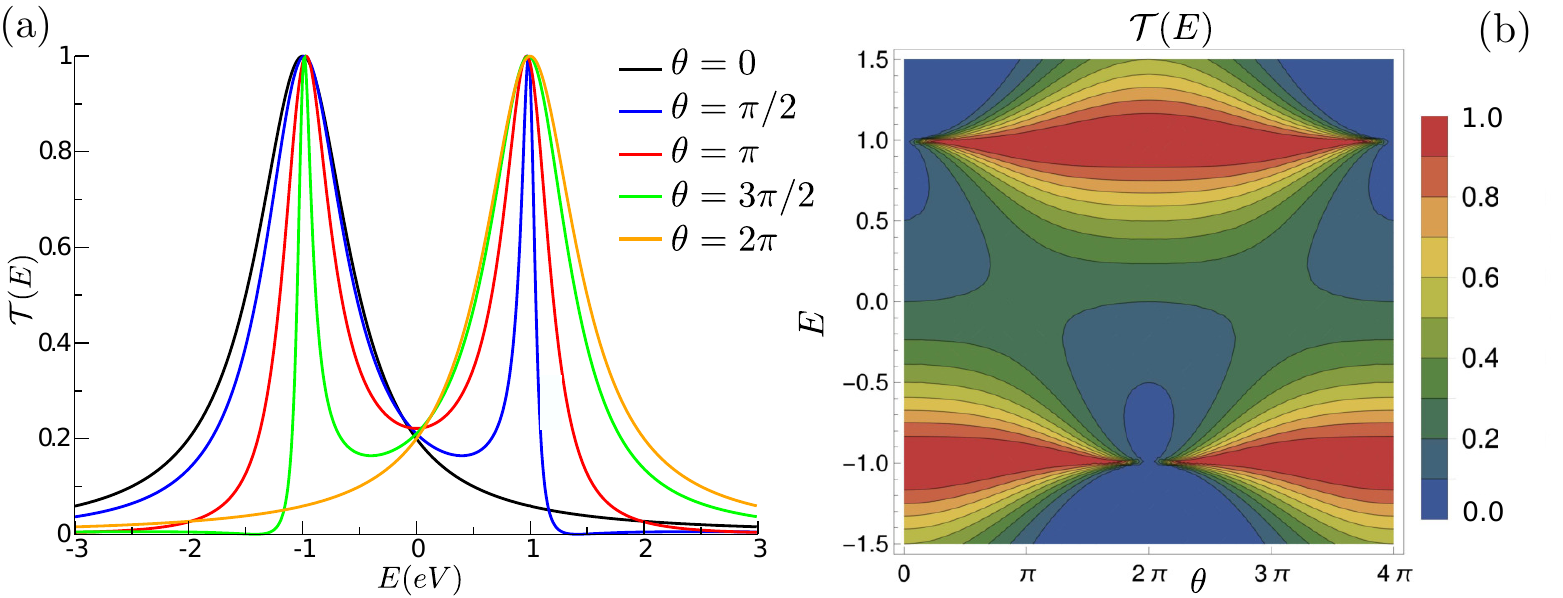}
\end{center}
\caption{ (a) Energy behavior  of ${\mathcal T}(E)$  versus $E$ for different   magnetic fluxes 
    $\theta = 2\pi\Phi(B)/\Phi_0$; 
    (b) contour plot  of ${\mathcal T}(E)$ as function of $\theta$ and $E$.
    The chosen parameters for the electronic system are  $E_d=0$ eV,  $t_d=-1$ eV and $\gamma= 0.25$ eV. 
    }
    \end{figure}

           \section{Structure of the transmission function and analytic evaluation of the kinetic parameters}

     Once the transmission function is known, we can access  the kinetic
     transport coefficients that control, in the linear approximation, the 
     thermoelectric properties of the  nanoscale  device.  The  kinetic 
     transport coefficients, in dimensionless form,  are linked to the 
     transmission function ${\mathcal T}(E)$  by the relations:
           \begin{equation} 
             K_n =   \int_{-\infty}^{+\infty} dE   \, {\mathcal T} (E) \, 
                               \frac{(E-\mu)^n}{(k_BT)^n}
                        \, ( -  \frac{ \partial f}{\partial  E} )   \qquad (n=0,1,2)   \ .
         \end{equation}  
   where $\mu$ is the chemical potential, $T$  the absolute temperature, 
   and $f(E,\mu,T)$ the Fermi function. 
   
   In the literature, the evaluation of the kinetic coefficients $K_{0,1,2}$ is 
   in general carried out either with the Sommerfeld expansion,\cite{ASHC} possibly extended up to fourth order,\cite{SILVA12} or by numerical integration.
     A nice aspect of the Sommerfeld expansion   is that the  procedure is analytic; however it holds only at 
  sufficiently low temperatures and reasonably smooth transmission function in the energy interval $k_BT$.
  The alternative procedure, based on numerical
  integration, requires particular caution because of the presence of sharp
  resonances and anti-resonance produced by the interference effects 
  of the magnetic fields. This is an obstacle to the construction of counter plots 
  or three dimensional graphics, often very useful  to  better 
  illustrate  at glance thermoelectric properties.
  The purpose of this section is to develop a brand new analytic procedure 
  for the evaluation   of the kinetic parameters, valid for any temperature range 
   and applicable in any desired domain of the other parameters at play.

  From  the structure  of Eq.(25) it is natural  to define 
  the {\it kinetic functional of order $n$} as follows
           \begin{equation}  
             {\mathcal F}_n[\ldots]  =  \int_{-\infty}^{+\infty} dE
                \,  [\ldots] \,  \frac{(E-\mu)^n}{(k_BT)^n}
                           \, ( -  \frac{ \partial f}{\partial  E} )  
                           \quad , \quad  f(E) = \frac{1}{e^{(E-\mu)/k_BT} + 1} \ ,
         \end{equation}  
   where $ [\ldots]$ stands for any arbitrary function of $E$ for which 
   the integral  exists. Then,  the  expression of the kinetic
   parameters  of the symmetric double dot  reads 
              \begin{equation}
                             K_n = {\mathcal F}_n [ {\mathcal T}(E) ] \ ,
              \end{equation} 
   where ${\mathcal T}(E)$ is the transmission function  reported in Eq.(18).  
   
   The first step to elaborate  analytically   the kinetic 
   functionals  requires the examination of  the pole 
   structure of ${\mathcal T}(E)$.  The transmission function can in fact 
   be resolved  into the sum of just  two simple poles, with appropriate 
   weighting factors. This is shown in detail in Appendix A.  
   According to  Eq.(A10),  the transmission function of the symmetric  
   double    dot can be cast in the form  
           \begin{equation} 
              {\mathcal T}(E) = 8 \gamma^2 \,  {\rm Re}  \left\{  \frac{1}{A_1} 
           \, \frac{ [\cos (\theta/2) \, (E\!-\!E_d)  +  t_d]^2} {E-z_1} 
                + \frac{1}{A_2} \, \frac{ [\cos (\theta/2) \, (E\!-\!E_d)  +  t_d]^2} 
                                 {E-z_2}    \right\}  
            \end{equation}
      where the pole positions $z_{1,2}$ and the weighting 
      factors $A_{12}$ are given by Eq.(A9).

              \begin{table} [t]
  \centering 
  {\begin{tabular}{c} \hline 
  Transmission function ${\mathcal T}(E)$ for the coupled degenerate double dot \vspace{3mm}  \\ \vspace{3mm}
         $
              {\mathcal T}(E) = 8 \gamma^2 \,  {\rm Re}  \left\{  \dfrac{1}{A_1} 
           \, \dfrac{ [\cos (\theta/2) \, (E\!-\!E_d)  +  t_d]^2} {E-z_1} 
                + \dfrac{1}{A_2} \, \dfrac{ [\cos (\theta/2) \, (E\!-\!E_d)  +  t_d]^2} 
                                 {E-z_2}    \right\}  
            $ \\ 
           where $ \quad
              \left\{     \begin{array}{l}
                     A_1     = -16\, i\gamma \cos^2(\theta/4) 
                           \, [- i\gamma \cos(\theta/2) + t_d] \, [- i\gamma + t_d]
                                                     \\[2mm]
                    A_2   = -16 \,i\gamma \sin^2(\theta/4)
                                \,  [ + i\gamma \cos(\theta/2) - t_d] \, [ -i\gamma - t_d] 
                                                  \\[2mm]
                    z_1 = E_d +t_d  -  i \gamma  -  i \gamma \cos(\theta/2) 
                                           \\[2mm] 
                   z_2 = E_d -t_d -  i \gamma  +  i \gamma \cos (\theta/2)   \ .
                       \end{array} \right.                                                              
         $\vspace{2mm} \\ 
                \hline
  {\begin{tabular}{c} \hline  \vspace{-7mm} \\
  Dimensionless kinetic parameters for the degenerate double dot system in the linear regime:\\ $  K_n =  \int dE  \, {\mathcal T} (E) \dfrac{(E-\mu)^n}{(k_BT)^n}  
                         ( -  \dfrac{ \partial f}{\partial  E} ) $
                        \hspace{1cm} \vspace{2mm}\\ \hline \vspace{2mm}
    $\begin{array}{l}
    \vspace{1mm}
             K_0   =   8 \gamma^2 \,  {\rm Re} \, \biggl\{ \dfrac{1}{A_1} 
                 \biggl[ \cos^2(\theta/2) \, (\mu + z_1 - 2E_d) +  2\cos(\theta/2) \, t_d  
                                        \biggr. \biggr.  \\
         \hspace{3cm}  +  \biggl. \biggl.     
                   \dfrac{ \left[\cos(\theta/2) \, (z_1 - E_d) + t_d\right]^2 }{k_BT}
                         \, I_{0}(\dfrac{z_1 - \mu}{k_BT})  \biggr] + \dfrac{1}{A_2}
                                     [ the \ same \ with \ z_2] \biggr\} 
                                           \\[1mm]
                K_1   =   8 \gamma^2 \,  {\rm Re} \, \biggl\{ \dfrac{1}{A_1} \biggl[
                \dfrac{\pi^2}{3}  \cos^2(\theta/2) \, k_BT  
                                        \biggr. \biggr.     \\
              \hspace{3cm} +  \biggl. \biggl.     
                   \dfrac{ \left[\cos(\theta/2) \, (z_1 - E_d) + t_d\right]^2 }{k_BT}
                         \, I_{1}(\dfrac{z_1 - \mu}{k_BT})  \biggr] + \dfrac{1}{A_2}
                                     [ the \ same \ with \ z_2] \biggr\}   
                                                 \\[1mm]
             K_2   =   8 \gamma^2 \,  {\rm Re} \, \biggl\{ \dfrac{1}{A_1} \biggl[
              \dfrac{\pi^2}{3}  \cos^2(\theta/2) \, (\mu + z_1 - 2E_d) 
              +  \dfrac{2\pi^2}{3}\cos(\theta/2) \, t_d  
                                        \biggr. \biggr.    \\
                 \hspace{3cm}  +  \biggl. \biggl.     
                   \dfrac{ \left[\cos(\theta/2) \, (z_1 - E_d) + t_d\right]^2 }{k_BT}
                         \, I_{2}(\dfrac{z_1 - \mu}{k_BT})  \biggr] + \dfrac{1}{A_2}
                                     [ the \ same \ with \ z_2] \biggr\}    
             \end{array} $
              \\ 
               $  I_{0}(w) =   \pm  \dfrac{1}{2\pi i} \Psi_t \, (\dfrac{1}{2} \pm \dfrac{i w}{2\pi})
                                   \ \   \,\,\, {\rm Im} \, w \lessgtr 0; \ \
                     I_{1}(w) = 1 + w I_{0}(w); \ \
                                I_{2}(w) = w + w^2 I_{0}(w) $ \vspace{2mm} \\                                                 
 \hline 
  \end{tabular}}
    \end{tabular}}
    \caption {{\footnotesize
Transmission function ${\mathcal T} (E)$ and 
        kinetic integrals $K_{0,1,2}$ in analytic form of the symmetric  double
     quantum dot, with two orbitals of the same diagonal  energy $E_d$, 
      coupled together by the off-diagonal hopping element $t_d$, in the wide 
      band approximation of parameter $\gamma$.  The phase $\theta$
      equals $2\pi \Phi(B)/\Phi_0$, where $\Phi(B)$ is the flux of magnetic field 
      through the nanodevice   in units of a single quantum flux $\Phi_0$. 
      The trigamma function is denoted with $\Psi_t$. } }
  \end{table}
             Equation (28) is fully equivalent to Eq.(18), but it enjoys the invaluable 
     advantage to put   in evidence its  two pole analytic structure. 
     This permits the  straight evaluation of the kinetic parameters: 
               \begin{eqnarray} 
                       K_n =  {\mathcal F}_n [ {\mathcal T}(E) ]  
                          &=&   8 \gamma^2 \,  {\rm Re} \, \left\{ 
                \frac{1}{A_1} {\mathcal F}_n \, \left[ \frac{ [\cos (\theta/2) 
                                                \cdot (E -E_d)  +  t_d]^2}  {E-z_1}  \right] \right.
                                                     \nonumber\\[1mm]
           & & \left.  \hspace{2cm}
                  + \frac{1}{A_2}  {\mathcal  F}_n \, \left[\frac{ [\cos (\theta/2) 
                               \cdot (E-E_d)  +  t_d]^2} {E-z_2} \right]   \right\}  .
            \end{eqnarray} 
      The analytic expressions of the kinetic functionals entering Eq.(29) 
       are provided in Appendix B.  The results for $K_0, K_1,K_2$
       are given by Eqs.(B11,B12,B13) respectively.   The transmission 
       function and the corresponding  kinetic integrals  of the symmetric 
       double dot   are summarized  in Table I,  for immediate reference. The procedure here outlined   is of value 
       not only for the present problem, but also because it
       provides useful guidelines for a number of more complex situations.
       
 \begin{table} [h]
  \centering 
 { \begin{tabular}{llr}
   \vspace{-2mm}\\ \hline
\multicolumn{3}{c}{  Expressions of the thermoelectric functions in terms of the kinetic parameters} \\  \hline
 \vspace{-3mm}\\
$\sigma =   K_0 \,  \dfrac{e^2}{h} $ & \hspace{1cm} $S = -   \dfrac{K_1}{K_0} \, \dfrac{k_B}{e}$ & $  \kappa_{e} = T ( K_2 - \dfrac{K_1^2}{K_0} )( \dfrac{ k_B^2}{h})  $ \vspace{2mm} \\   
$\dfrac{ {\mathcal P} }{\eta_c^2 }= \dfrac{1}{4} \, T^2 \, \dfrac{K_1^2}{K_0}  \, \dfrac{ k_B^2}{h}   $ & \hspace{1cm} $L = \dfrac{ \ K_0 K_2 - K_1^2 \ } {K_0^2} \  \dfrac{ k_B^2}{e^2}$  \vspace{3mm}    &\hspace{1cm} $p =  \dfrac{K_1^2}{K_0K_2}  \quad  (0\le p \le 1)$  \vspace{2mm}\\ 
\multicolumn{3}{l}
{$ZT=\dfrac{K_1^2}{K_0 K_2 - K_1^2}=\dfrac{p}{1-p} $  \hspace{4cm}        $\dfrac{\eta}{\eta_c} = \dfrac{\sqrt{1+ZT}-1} {\sqrt{1+ZT}+1}=    \dfrac{1-  \sqrt{1-p} } { 1+ \sqrt{ 1 - p}} $} \vspace{2mm}\\
    \hline
\multicolumn{3}{c}{  Expressions of the thermoelectric natural units  for nanoscale devices} \   \vspace{2mm}\\ \hline
     $\dfrac{e^2}{h}  =  3.874046 \cdot 10^{-5} \  \dfrac{\rm A}{\rm V} $\ ; \,\,& \hspace{1cm} $ \dfrac{k_B}{e}  =   86.17 \  \dfrac{ \mu {\rm V} }{\rm K}$  \ ; \,\,& $\dfrac{k_B^2}{e^2}   = 74.25 \cdot 10^{-10} \, \dfrac{ {\rm V}^2}{ {\rm K}^2} $ ; \vspace{2mm}\\
     \multicolumn{3}{l}
{     $\dfrac{k_B^2}{h}  = 1.8 \cdot 10^{6} \, \dfrac{\rm eV}{\rm sec} \cdot \dfrac{1}{\rm K^2} =0.288\,\dfrac{\rm pW}{\rm K^2} $
$ \quad  \Longrightarrow   \dfrac{k_B^2T_0^2}{h}  = 1.8 \cdot 10^{6} \, \dfrac{\rm eV}{\rm sec} =0.288 \,{\rm pW} \quad (T_0=1K) $}
     \vspace{2mm}\\
         \hline
    \end{tabular}} 
  \caption{ {
  Transport parameters  in the linear approximation  
       for  thermoelectric materials, with  electronic transmission 
       function ${\mathcal T}(E)$.  
     The kinetic parameters $K_{0,1,2}$ are 
      defined   in dimensionless form.    The  electric conductance $\sigma$, 
      Seebeck coefficient $S$, power-output $ {\mathcal P}$,  electronic thermal conductance $\kappa_e$, 
     Lorenz number $L$, performance parameter $p$, figure of merit  $ZT$ and efficiency ${\eta}$ 
     are reported. The quantity  $\eta_c$ denotes 
     the Carnot  efficiency $\eta_c  = \Delta T/T$, where $\Delta T$ is 
     the temperature difference between the hot reservoir and 
     the cold one.  } }
  {}
  \end{table}
    
 After achieving the task of a straight analytic evaluation of the kinetic parameters of the double quantum dot system as summarized in Table I, it becomes now routine to investigate the transport properties. Following closely Ref.~\onlinecite{NOI}, in Table II we report for sake of completeness the expressions of the electric and thermal conductances, of the Seebeck coefficient and the other transport parameters of interest, in terms of the kinetic coefficients $K_{0},K_{1,}$, and  $K_{2}$.

     In the next section we evaluate  magneto transport properties of specific double dot devices, and discuss the variety and wealth of effects occurring in spite of the reasonable simplicity of the model.

      

                 \section{Results and discussion} 
      
      We begin to examine a realistic space domain for the thermoelectric device under attention.
    For molecular junctions, we can 
     set $\gamma\approx 0.25 \ {\rm eV}$ and $t_d \approx -1.0 \ {\rm eV}$.
     The fact that $|t_d|>> \gamma$ (almost an order of magnitude) assures
     that in the transmission function the Lorentz lineshape and the Fano lineshape are in general well
     resolved,  with linewidths  $2\gamma \cos^2(\theta/4)$ 
     and $2\gamma \sin^2(\theta/4)$, respectively, as it is seen 
     from Eq.(20). The values of $\theta$  explored  to better highlight periodicity  as function of $\theta$, are in the 
     whole range $[0,4\pi]$, and in particular   $\theta=0, \pi/2, \pi , 3\pi/2$ and $2\pi$. 
     The range of  $\theta$ from one flux to two flux quanta ($2\pi<\theta\leq4\pi$) retraces back the range from one flux to zero, and does not need to be considered explicitly. The
     room temperature  considered entails $k_BT=0.025 \ {\rm eV}$.     The dot energy $E_d$ is taken as the reference energy 
     and set equal to zero. In summary,: the figures reported below in this section refer to the set of parameters $E_d=0$, $\gamma=0.25$ eV, $t_d=-1.0$ eV, $k_BT=0.025 \ {\rm eV}$  and $\theta=0, \pi/2, \pi , 3\pi/2$ and $2\pi$.  When useful, other temperatures, phases or parameter domain have been explored and commented (but in general not explicitly reported).
     
     In Fig.3  the thermoelectric functions of the c-2QD, for varying chemical potential $\mu$ and magnetic flux parameter $\theta$ are provided. The left panels show the landscape  of electrical conductivity $\sigma$, electrical thermal conductivity $\kappa_e$, Seebeck coefficient $S$,  and figure of merit $ZT$.  The right panels  show sections of the same quantities for -2 eV$<\mu<$2 eV at $\theta=0,\pi/2,\pi, 3\pi/2$  and  $2\pi$, to better highlight their shape and symmetry.
     The curves profiles reported in the left panels  respect the color sequence  shown in the corresponding right panels.
     From Fig 3a and Fig.3b we observe that $\sigma$ and $\kappa_e$  have behavior similar to ${\mathcal T}$(E), as expected from their expressions; we also verified that the value of $\sigma$ increases 
(not shown in the figures) decreasing the temperature, and that the opposite  occurs for $\kappa_e$. 
   \begin{figure}[h!]
\begin{center}
\includegraphics[width=0.76\linewidth]{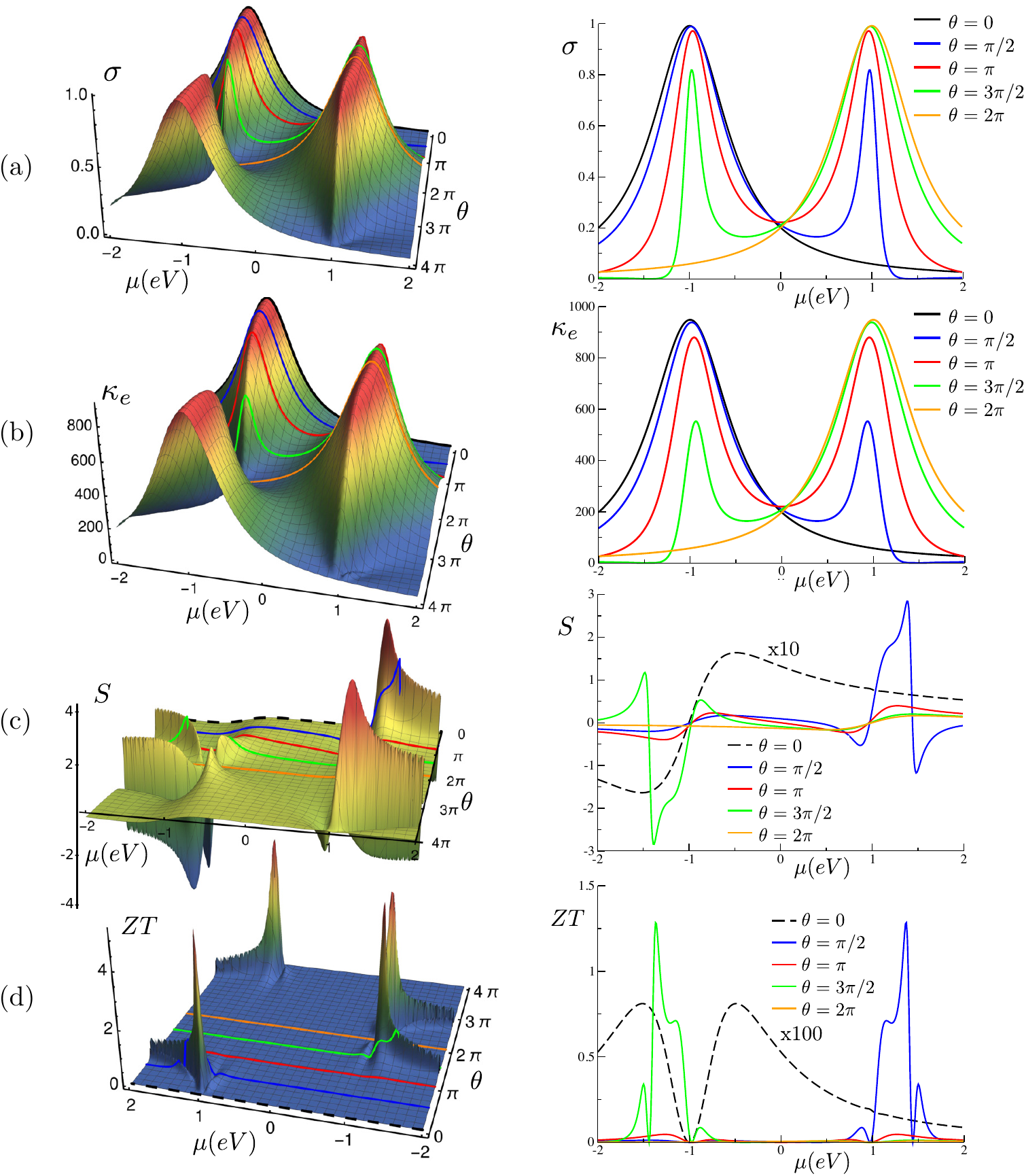}
\end{center}
\caption{{\small
 (a)  {{  Electrical conductivity $\sigma$ (in units ${e^2}/{h}$).  (b) Electrical thermal conductivity $\kappa_e$ (in units ${k_B^2}T_0^2/{h}$). (c) Seebeck coefficient $S$ (in units ${k_B}/{e} $).  (d) Figure of merit $ZT$ of the c-2QD under attention in the ($\mu-\theta  $) plane.} } The left panels report the landscape of the thermoelectric functions  in the ($\mu-\theta$) plane,   the right panels   report  sections  of the same quantities  at $\theta=0,\pi/2,\pi, 3\pi/2$  and  $2\pi$.
 The black dashed lines in the right (c) and (d) panels evidence the results in the absence of magnetic flux.  For $\theta=2\pi$ no multiplication by 10 or by 100 has been performed, to better emphasize the enhancement effect  of the magnetic field.
 The colored curves  in the left panels  respect the  sequence  of the graphs shown in the corresponding right panels.}
 }
    \end{figure}
      
      We observe that in the absence  of magnetic field,  i.e. $\theta$=0, ${\mathcal T}$(E) presents a Breit-Wigner resonance
 around $E_b=-1$ eV, and similarly   $\sigma(\mu,0)$, and $\kappa_e(\mu,0)$ present a Breit-Wigner resonance
 around $\mu=E_b$. 
Moreover, near the resonant energy  the thermopower $S$ vanishes while for $\mu\lesssim E_b$ ($\mu\gtrsim E_b$) $S$ is negative (positive), indicating  mainly $n$-type ($p$-type) behavior of the device.
The figure of merit $ZT$ vanishes   where $S$ vanishes  as expected from  its definition, and remains small ($<0.01$)  for any $\mu$. As temperature increases  both $S$ and $ZT$ values increase.
 
     When the magnetic field is switched on, both Breit-Wigner- and Fano-like resonances may contribute to the transmission spectra.  In particular, for $\theta=2n\pi$, with $n$  integer  number, only  Breit-Wigner resonances occur, which are located  at the bonding energy for $n$ even and at the antibonding energy for $n$ odd [see Eq. (21) and Eq. (22)]. For $\theta=(\pi/2+n\pi)$ both Breit-Wigner- and Fano-like resonances are present in the ${\mathcal T}$(E) spectrum, with Breit-Wigner (Fano)  features centered at the bonding (antibonding) energies  for $n$ even and viceversa for $n$ odd. We notice that ${\mathcal T}$(E)  is symmetric around $E_d$ for $\theta=\pi$ or $\theta= (2n+1)\pi$ as required by Eq. (23).
It is important to  observe that $|S|$ increases by more than 10 times and $ZT$ by more than 100 times  with respect to the case $\theta=0$, for specific values of the magnetic flux threading the c-2QD circuit as evidenced in the plots in the right side of Fig.3c and Fig.3d. In particular $|S|$ assumes large values ($\approx4$ k$_B$/e) in the regions around $\theta \sim \pi/2$ and $\theta \sim 3\pi/2$ in the resonance and in the antiresonance regions.
 The above results are   in agreement with  the ones obtained  for the benzene molecule junction in magnetic flux.\cite{LI17}
 Fig.3d shows that  for the chosen $T$ and $\gamma$ parameters,   $ZT$ can reach  values  $\approx 6$ in the regions  $\theta \sim \pi/2$ and $\theta \sim 3\pi/2$. The above results   evidence that  temperature and magnetic flux can be exploited to increase the thermoelectric factor   of merit .
 
   \begin{figure}[h!]
\begin{center}
\includegraphics[width=0.84\linewidth]{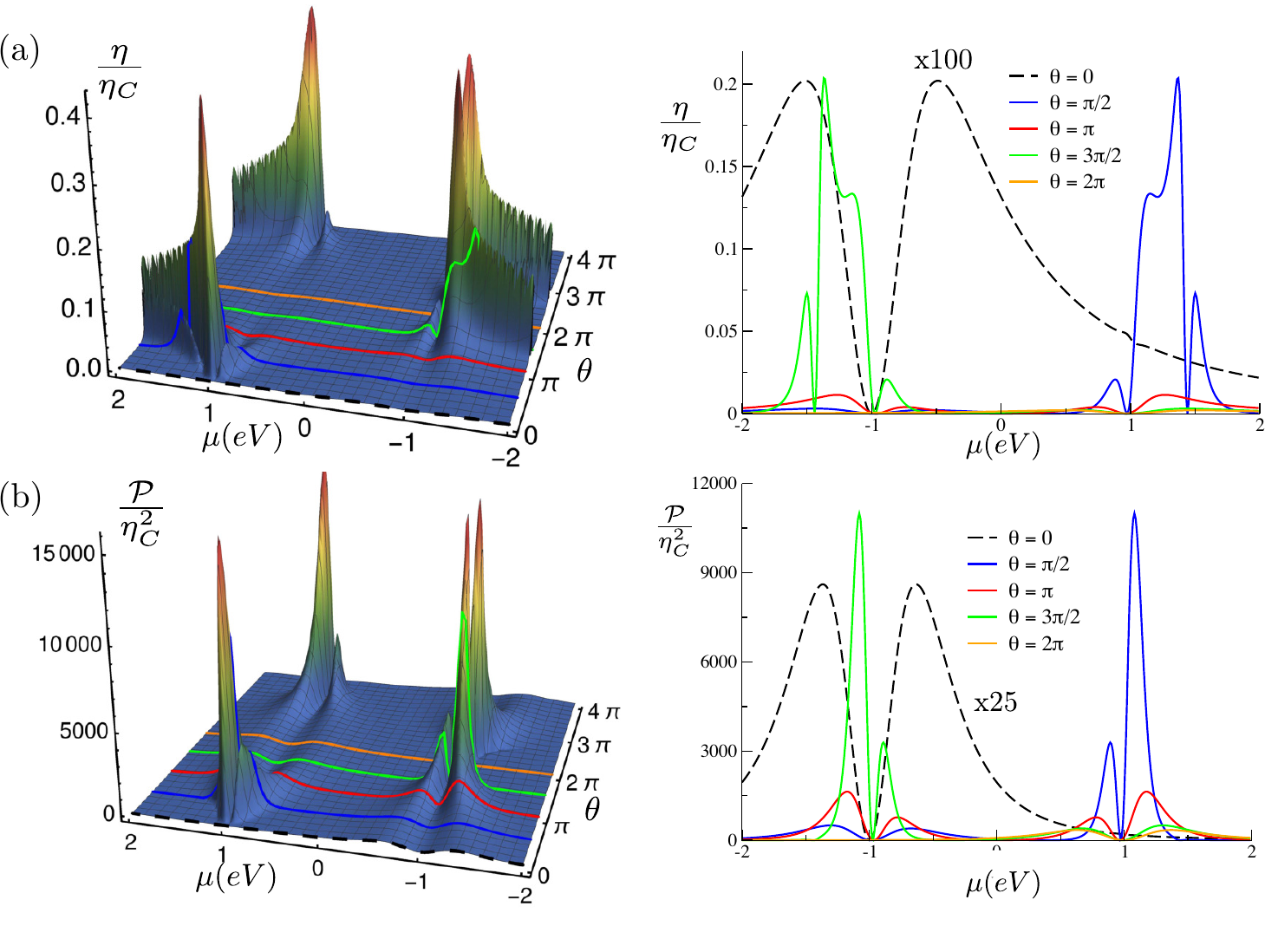}
\end{center}
\caption{  {{ (a) Thermoelectric efficiency $\eta/\eta_c$ in the ($\mu-\theta$) plane. b) Output power  ${\mathcal P}/\eta_c^2$ (in units $k_B^2T_0^2/h$).}}
The left panels of Fig.4a and Fig.4b report the landscape of the thermoelectric efficiency and power output   in the ($\mu-\theta$) plane,   the right panels   report  sections  of the same quantities  at $\theta=0,\pi/2,\pi, 3\pi/2$  and  $2\pi$. The black dashed lines in the right   panels evidence the results in the absence of magnetic flux.  }
    \end{figure}
    

     Most interesting is the evaluation of the performance  of the c-2QD as heat engine, in this case  a study of the  efficiency at the maximum output power is required. Several recent papers~\cite{LINKE05,NAP10,ESP10,WIT13,HER13,LUO16} 
     have shown that the mere knowledge of the maximum efficiency of a heat engine is of limited importance since the useful operative information  concerns   the conditions corresponding to the maximum power output.\cite{CA75,CASATI17} It is known in fact,that even if the  figure of merit $ZT$ of a thermoelectric device can assume large values ($>>$1) mainly  for nanostructured systems\cite{WIERB11,VAR01,MUR08}, what really matters  is just the efficiency evaluated at the maximum power output.
     To better clarify this point, we report in Fig.4a the thermoelectric efficiency  and  in Fig.4b  the output power, respectively,  as function of $\mu$ and $\theta$, as defined in Table II.

Once again  we observe that the magnetic field strongly enhances the thermoelectric efficiency by more than two orders of magnitude with respect to the case of absence of magnetic field, in the resonance and antiresonance regions, while output power increases  more than 25 times and can assume values of the order of $10^4$  {{  (in units $k_B^2T_0^2/h$)}}.
Fig.5  summarizes the results of the evaluation of the   efficiency at the maximum power output,  which is the  most appropriate metric to measure the performance of the device.
For this aim we have scanned the flux $\theta$ parameter in the [0-4$\pi$] range and, for any $\theta$, we have looked for  the    maximum  output power for varying values of  the chemical potential $\mu$.
This has allowed to evaluate   the efficiency for the  values of $\theta$ and $\mu$ which determine the maximum power conditions.
     
      \begin{figure}[t]
\begin{center}
\includegraphics{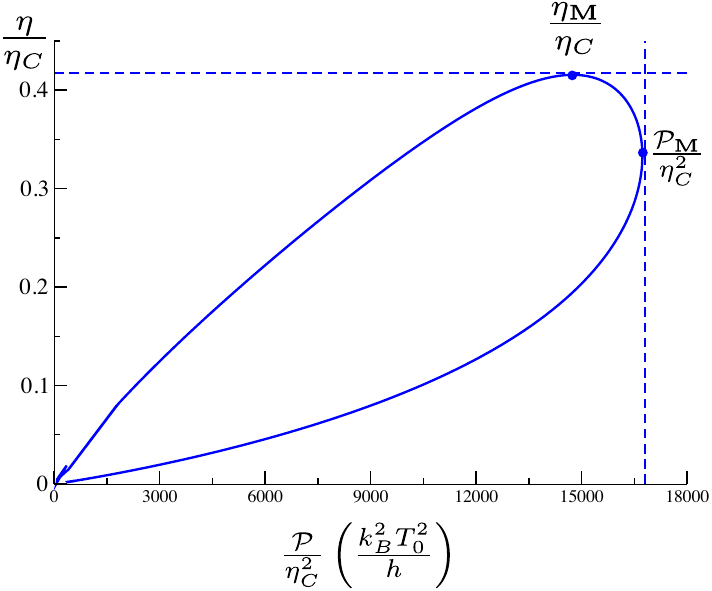}
\end{center}
\caption{Maximum efficiency at the maximum output power. ${\mathcal P}_M/\eta_C^2$ is the highest value of the output power; $\eta_{M}/\eta_C$  is the maximum value of the device efficiency.}
    \end{figure}

     The set of all the maximum power and corresponding efficiency data have been exploited to produce  Fig.5 which reports the curve of the  maximum efficiency at the maximum output power.  From Fig.5 we can observe that the maximum efficiency is higher than the efficiency at operating conditions where the maximum output power is realized
   We can see that  the highest value of the power output    ${\mathcal P}_M/\eta_C^2$ is 16800  {{  (in units $k_B^2T_0^2/h$)}}
     for the values
      $\theta\approx 1$, and $\theta\approx(4\pi-1)$, at  $\mu\approx1.062$ eV,  and for 
     $\theta\approx(2\pi-1)$ and  $\theta\approx(2\pi+1)$, at  $\mu\approx-1.062$ eV. 
      Correspondingly,  the  normalized efficiency  at maximum power is  $\eta({\mathcal P}_M)/\eta_C$=0.33. 
     Moreover, we can see that the highest value of efficiency    $\eta_{M}/\eta_C$  is 0.43  which occurs for the values  $\theta\approx\pi/4$ and   $15\pi/4$, at  $\mu\approx1.068$ eV and for 
     $\theta\approx7\pi/4$ and  $9\pi/4$, at  $\mu\approx-1.068$ eV. Correspondingly, the power output is   
     ${\mathcal P}(\eta_M)/\eta_C^2= 14200$  {{   (in units $k_B^2T_0^2/h$)}}.


      \section{Conclusions} 
      
      We have presented in this paper  a systematic analytic study of the thermoelectric response functions of a coupled double quantum dot system, pierced by a magnetic field, connected to left and right reservoirs, in the linear regime.
    Our method is based on the Green$'$s function formalism. The  results   are analytic and can be expressed  in terms of  easily accessible trigamma functions and Bernoulli numbers; this has allowed  to scan  wide ranges of  values of  chemical potentials and temperatures of  the reservoirs, different threading magnetic fluxes, dot energies and interdot interactions.
Our results  show that  thermoelectric transport through the c-2QD can be strongly enhanced  by the magnetic flux, mainly in the energy regions around the bonding and antibonding resonances of the system, which can be experimentally reached varying the system chemical potential by appropriate gate. The thermopower $S$ can be  enhanced by more than ten times    and the figure of merit  $ZT$ by more than hundred times by the presence of a threading magnetic field. Most important, we have also found in this simple system that  the magnetic flux increases  the performance  of the device under maximum power output conditions.

    \vspace{1cm} 
             
     
  \appendix
     
        \vspace{0.5cm}
\noindent {\bf {\large Appendix A. Transforming product of simple 
      poles into the weighted sum of simple poles}}	
 \vspace{0.5cm}    
            
                  \setcounter{equation} {0}
       \numberwithin{equation}{section}
       \renewcommand\theequation{A\arabic{equation}}
  
    The purpose of this Appendix is to  transform a product of simple poles
     into the fully equivalent  (and much more convenient)  weighted sum of 
     simple poles. Without entering into technicalities and subtleties, 
     we confine our attention to the case of four poles, just of actual interest 
     for double dots. We start from the identity
             \begin{equation}
              \frac{1}{(E \!-\! z_1)(E  \!-\! z_2)(E \!-\! z_3)(E \!-\! z_4)}  \equiv 
      \frac{1}{A_1} \frac{1}{E \!-\! z_1} \!+\!  \frac{1}{A_2}\frac{1}{E \!-\! z_2} 
         \!+\! \frac{1}{A_3}\frac{1}{E \!-\! z_3} \!+\!  \frac{1}{A_4} \frac{1}{E \!-\! z_4}
      \end{equation}     
    where the $A_{1,2,3,4}$ constants (i.e. independent from the energy 
    variable)  have the expressions
          \begin{eqnarray}
                \left \{      \begin{array}{lll}
                        A_1 &=&   (z_1 -z_2) (z_1 -z_3) (z_1 -z_4)    
                                          \\[1mm]
                       A_2 &=&   (z_2 -z_1) (z_2 -z_3) (z_2 -z_4)  
                                         \\[1mm]
                      A_3 &=&   (z_3 -z_1) (z_3 -z_2) (z_3 -z_4) 
                                            \\[1mm]
                       A_4 &=&   (z_4 -z_1) (z_4 -z_2) (z_4 -z_3)  
                              \end{array} \right.   \ .                                                            
        \end{eqnarray}  
      It is seen that the $A_i$ constant is the product of the differences of $z_i$
      with all the other poles $z_j (\neq z_i)$, except $z_i$ itself.
      
      To demonstrate the identity (A1), suppose to multiply both 
      members of Eq.(A1) by 
      the expression $\Pi_{i=1,4}(E-z_i)$. After multiplication, the  first 
      member becomes independent from $E$, and equal  to  unit. 
      This occurs also for the second 
      member. In fact after multiplication, the second member becomes a 
      polynomial in $E$ of order three,  which takes the unity value
      at the four arguments $E=z_1,z_2,z_3,z_4$, and is thus   unity everywhere.

     \vspace{1cm}
\noindent {\bf Case of complex conjugate poles}	
 \vspace{0.5cm}

       In the case of complex  conjugate poles, 
       say $(z_1 ,  z_2  ,  z_3 = z_1^\ast , \  z_4 = z_2^\ast )$,  it is seen 
       by inspection that Eqs.(A1,A2) simplify in the form
           \begin{equation}
              \frac{1}{(E - z_1)(E  - z_2)(E - z_1^\ast)(E - z_2^\ast)}  
              \equiv   2 \, {\rm Re} \, \left[ 
           \frac{1}{A_1} \frac{1}{E-z_1} +  \frac{1}{A_2}\frac{1}{E-z_2} \right]  ,
      \end{equation}     
         where the $A_{1,2}$ constants  have the expressions
          \begin{eqnarray}
                \left \{      \begin{array}{lll}
                        A_1 &=&   (z_1 -z_2) (z_1 -z_1^\ast) (z_1 -z_2^\ast)    
                                          \\[1mm]
                       A_2 &=&   (z_2 -z_1) (z_2 -z_1^\ast) (z_2 -z_2^\ast)    
                              \end{array} \right.   .                                                            
        \end{eqnarray}  

\vspace{0.5cm}
\noindent {\bf Pole structure of the double dot transmission function}
 \vspace{0.5cm}

        The expression of the transmission function of the symmetric double 
        dot is provided in Eq.(18), and can be written in the form 
           \begin{equation} 
              {\mathcal T}(E) = \frac{4 \gamma^2}{D^R(E)D^A(E)}
                      \left[  \cos (\theta/2) \cdot (E-E_d)  +  t_d  \right]^2  
            \end{equation} 
    where  $D^A(E) = [D^R(E)]^*$, and
           \begin{equation} 
                   D^{R}(E) =  [E  - E_d  -  t_d  
                            +  i \gamma  + i\gamma \cos (\theta/2)] \, 
	             [E  - E_d  + t_d  +  i \gamma  - i\gamma \cos (\theta/2)] \ .
	    \end{equation}  

     The analytic structure of the  transmission function can be put
      in better evidence by expressing  $D^{R}(E)$  in the form
           \begin{equation}
                D^{R}(E)= (E - z_1)(E- z_2)  \qquad \left\{
                             \begin{array}{l} 
                              z_1 = E_d +t_d -  i \gamma  -  i \gamma \cos(\theta/2) 
                                            \\ [2mm] 
                            z_2 = E_d - t_d -  i \gamma  +  i \gamma \cos (\theta/2) 
                             \end{array} \right.  \ , 
                \end{equation}
          and similarly for its complex conjugate $D^{A}(E)$.
      Using Eqs.(A3,A4,A7) one obtains 
          \begin{equation}
              \frac{1}{D^R(E) D^A(E)}  
              \equiv   2 \, {\rm Re} \, \left[ 
           \frac{1}{A_1} \frac{1}{E-z_1} +  \frac{1}{A_2}\frac{1}{E-z_2} \right]\ ,
      \end{equation}     
           where
          \begin{equation}
              \left\{     \begin{array}{rcl}
                     A_1     &=& -16\, i\gamma \cos^2(\theta/4) 
                           \, [ \,\, t_d - i\gamma \cos(\theta/2) \,\, ] \, [\,\, t_d - i\gamma \,\, ]
                                                     \\[2mm]
                    A_2   &=& -16 \,i\gamma \sin^2(\theta/4)
                                \,  [ -t_d + i\gamma \cos(\theta/2) ] \, [-t_d  -i\gamma ] 
                                                  \\[2mm]
                    z_1 &=& E_d +t_d  -  i \gamma  -  i \gamma \cos(\theta/2) 
                                           \\[2mm] 
                   z_2 &=& E_d -t_d -  i \gamma  +  i \gamma \cos (\theta/2)  \ .
                       \end{array} \right.                                                              
               \end{equation} 
      The transmission function of the symmetric  double can be expressed
       in the form  
           \begin{equation} 
              {\mathcal T}(E) \!=\! 8 \gamma^2 \,  {\rm Re}  \left\{  \frac{1}{A_1} 
           \, \frac{ [\cos (\theta/2) \, (E\!-\!E_d)  +  t_d]^2} {E-z_1} 
                + \frac{1}{A_2} \, \frac{ [\cos (\theta/2) \, (E\!-\!E_d)  +  t_d]^2} 
                                 {E-z_2}    \right\}  
            \end{equation}
        Eq.(A10) shows explicitly the pole structure of the transmission function, 
       and is ready for analytic evaluation of the corresponding kinetic integrals.

     

     \vspace{1cm}
\noindent {{\large \bf Appendix B. Kinetic functional  for the analytic 
               treatment of thermoelectricity in double dots}}	
 \vspace{0.5cm}      
       
      \setcounter{equation} {0}
       \numberwithin{equation}{section}
       \renewcommand\theequation{B\arabic{equation}}

      For the analytic treatment  of thermoelectricity in double dots, 
      it is convenient to define   the kinetic functional as follows
               \begin{equation}
                    {\mathcal F}_n [ \ldots] =  \int_{-\infty}^{+\infty} dE  
                 \, [ \ldots]
                   \, \frac{(E-\mu)^n}{(k_BT)^n} \, ( -  \frac{ \partial f}{\partial  E} ) 
                       \qquad    f(E) = \frac{1 }{e^{(E-\mu)/k_BT} + 1} \ ,
              \end{equation} 
       where $[ \ldots]$ stands for any arbitrary function of $E$, for which 
       the integral in  Eq.(B1) exists, and $n=0,1,2,...$.  What is needed 
       are just the  kinetic functionals  of  polynomials in energy,  and their
       product by a simple pole.  The results are of interest not only  
       in the model nanostructure under attention, but also in other more  
       general and complex models.

        \vspace{1cm}
\noindent {\bf Kinetic functional of a constant and of the 
                                variable itself}	
 \vspace{0.5cm}

       Due to the fact that the functional (B1) is linear,  the functional of a 
       constant equals the functional of unity times the constant. The 
       functional of unity is
              \begin{eqnarray} 
                      {\mathcal F}_n \left[1\right]  
                 &=&  \int_{-\infty}^{+\infty} dE  
                   \, \frac{(E-\mu)^n}{(k_BT)^n} \, ( -  \frac{ \partial f}{\partial  E} ) 
                       \qquad    f(E) = \frac{1 }{e^{(E-\mu)/k_BT} + 1}
                                             \nonumber    \\[2mm]
                  &=&   \int_{-\infty}^{+\infty}    dE       
	            \, \frac{(E-\mu)^n}{(k_BT)^n} \,  \frac{1}{k_BT}
                   \frac{  e^{  (E-\mu)/k_BT} }{ [e^{(E-\mu)/k_BT} + 1]^2} 
                   \quad ; \quad  set \quad   z = \frac{E-\mu}{k_BT} \ ;
                                            \nonumber        \\[2mm]
                  &=&    \int_{-\infty}^{+\infty}    dz \, z^n   \,  
                \frac{  e^z} { (e^z + 1)^2}  \equiv b_n   \ .
         \end{eqnarray}  
     Thus the kinetic functionals of unity  equal the Bernoulli-like numbers.
     The first few Bernoulli-like numbers $b_n$ are  
            \begin{equation}
                 b_0 = 1 \ , \ b_1 = 0 \ , \ b_2 = \frac{\pi^2}{3} \ , \ b_3 = 0 \ ,
                          \   b_4 = \frac{7 \pi^4}{15} \ , \  b_5 = 0  \ ,
                           \ b_6 = \frac{31 \pi^6}{21} \ \ldots\
             \end{equation}  
      [For details see Ref.\onlinecite{GRAD}].

     The kinetic functional of the energy is easily obtained, using again 
      the  linear properties of the functional. It holds
         \begin{equation*}
              {\mathcal F}_n \left[E - E_d\right] 
              =   {\mathcal F}_n \left[(E-\mu) + \mu - E_d\right]
                  =  k_BT\, {\mathcal F}_n \left[ \frac{E-\mu}{k_BT} \right]
                                  + (\mu - E_d) \, {\mathcal F}_n[1] \ .
            \end{equation*}
        It follows
              \begin{equation}
                    {\mathcal F}_n \left[E - E_d\right]   = k_BT  \, b_{n+1} 
                                            + (\mu- E_d)\, b_n \ .
             \end{equation} 
                  
  \vspace{1cm}
\noindent {\bf Kinetic functional of a simple pole}	
 \vspace{0.5cm}

      Consider the simple pole function of the form
           \begin{equation} 
                       X_0(E) = \frac{1}  { E- z_p}  \ ,
           \end{equation}  
     where  $z_p$ is  a given position  in the upper or lower part of 
     complex plane. The  kinetic functional  becomes
           \begin{equation*} 
                      {\mathcal F}_n \left[\frac{1}  { E- z_p}\right]  
                 =   \int_{-\infty}^{+\infty}    dE       
                   \,   \frac{ 1 } {E - z_p}
	            \, \frac{(E-\mu)^n}{(k_BT)^n} \,  \frac{1}{k_BT}
                   \frac{  e^{  (E-\mu)/k_BT} }{ [e^{(E-\mu)/k_BT} + 1]^2} \ .
         \end{equation*}      
    As usual, it is convenient to introduce the  dimensionless variables
           \begin{equation*}
            z = \frac{E-\mu}{k_BT}   \quad  ; \quad dz =  \frac{dE}{k_BT} 
                  \quad ; \quad   E = k_BT \, z + \mu \ .
         \end{equation*}        
     With the indicated substitutions, one obtains
             \begin{eqnarray*}
                {\mathcal F}_n \left[\frac{1}  { E- z_p}\right] 
                    &=&    \int_{-\infty}^{+\infty}    dz \, \frac{ z^n } 
                       {\, k_BT \, z + \mu - z_p \,}   \,  
                \frac{  e^z} { (e^z + 1)^2} 
	                              \nonumber \\[2mm]
	            &=& \frac{1}{k_BT}  \int_{-\infty}^{+\infty}   
	            dz   \,  \frac{ z^n } 
                  { z- (z_p - \mu)/ k_BT } 
	         \, ( -  \frac{ \partial f}{\partial  z} )   \quad  \ 
	                             f(z) = \frac{1}{e^{z} +1} 
	                                \nonumber \\[2mm]
	           &=& \frac{1}{k_BT}  \int_{-\infty}^{+\infty}    dz   \,  
	            \frac{ z^n } { z-  w_p }  \,  ( - \frac{ \partial f}{\partial  z} ) 
	               \qquad \quad  w_p = \frac{z_p - \mu}{k_BT}   \ .
         \end{eqnarray*}  
     In summary, it holds  
           \begin{equation} 
                  {\mathcal F}_n \left[\frac{1}  { E- z_p}\right]  
                \equiv  \frac{1}{k_BT} \, J_n(w_p)   
                                \qquad , \qquad  w_p = \frac{z_p - \mu}{k_BT}  \ ,   
         \end{equation}   
      where    $J_n$   denote the complex functions
           \begin{equation*}
             J_n(w) =  \int_{-\infty}^{+\infty}    dz   \,  
	            \frac{ z^n } { z-  w }  \,  ( - \frac{ \partial f}{\partial  z} )
	            \qquad n=0,1,2, \ldots  
	     \end{equation*}  
	 The first few  $J$-functions are
	       \begin{equation*}
                 J_{0}(w) =   \pm  \frac{1}{2\pi i} \Psi_t \, (\frac{1}{2} \pm \frac{i w}{2\pi})
                  \ \ {\rm Im} \, w  \lessgtr 0
                                    \ \ ; \ \
                     J_{1}(w) = 1 + w J_{0}(w) \ \ ; \ \
                                J_{2}(w) = w + w^2 J_{0}(w) \ ,
          \end{equation*} 
        where
          \begin{equation*} 
                    \Psi_{t}(z) = \sum_{n=0}^{\infty} 
                                  \, \frac{1}{(z+ n)^{2}}  
           \end{equation*} 
       is the trigamma function.
        The trigamma function is a one-valued analytic function with poles of order two at the points $ z=0, -1, -2, ...$, and routinely available in computer libraries. For details on the digamma, trigamma and 
       poligamma functions see Ref.\onlinecite{ABRA}

      Trigamma function and Bernoulli-like numbers are the  
        ingredients for the analytic evaluation of the kinetic functional 
        of interest, including the next ones.

  \vspace{0.8cm}
\noindent {\bf Kinetic functional of a simple pole times the first 
                  and second power of the energy}	
 \vspace{0.2cm}

      Consider the  function of the form
           \begin{equation*} 
                       X_1(E) = \frac{E - E_d}  { E- z_p}  \ ,
           \end{equation*} 
     where  $z_p$ is  the position of the pole in the upper or lower part of 
     complex plane, and $E_d$ is an arbitrary complex constant. We have
             \begin{equation*} 
                  {\mathcal F}_n \left[\frac{E - E_d}  { E- z_p}\right]
                       = {\mathcal F}_n \left[\frac{(E-z_p)+z_p - E_d}  { E- z_p}\right]
                       = {\mathcal F}_n \left[1\right] 
                          +(z_p-E_d)\,{\mathcal F}_n \left[\frac{1}  { E- z_p}\right]  .
              \end{equation*}
       It follows 
           \begin{equation} 
                  {\mathcal F}_n \left[\frac{E - E_d}  { E- z_p}\right] = b_n   
                    + \frac{z_p - E_d}{k_BT}\,I_n( w_p)  
                             \qquad ; \qquad  w_p = \frac{z_p - \mu}{k_BT} \ .
              \end{equation}  

      \noindent Another function to consider is
              \begin{equation*} 
                       X_2(E) = \frac{(E-E_d)^2}  { E- z_p}  \ ,
           \end{equation*} 
     where  $z_p$ is  a given position  in the upper or lower part of 
     complex plane. The  kinetic functional  can be cast in the form
          \begin{eqnarray*} 
                  {\mathcal F}_n \left[\frac{(E-E_d)^2}  { E- z_p}\right] 
                  &=&   {\mathcal F}_n \left[  E+ z_p -2E_d 
                                     + \frac{(z_p-E_d)^2}  { E- z_p}\right]
                                                 \\[1mm]
                  &=&  {\mathcal F}_n \left[ E - E_d\right] 
                          + (z_p - E_d) \, {\mathcal F}_n\left[ 1\right]
                       + (z_p - E_d)^2 \, {\mathcal F}_n \left[\frac{1}  { E- z_p}\right]  .
       \end{eqnarray*}           
    Using previous results we obtain
          \begin{equation} 
                  {\mathcal F}_n \left[\frac{(E-E_d)^2}  { E- z_p}\right] 
                   = k_BT\, b_{n+1} + (\mu + z_p - 2E_d)\, b_n    
                    + \frac{(z_p-E_d)^2}{k_BT}\,I_n( w_p)   \ .
       \end{equation}  
     We could proceed with  higher powers along similar lines, 
     whenever needed.

  \vspace{1cm}
\noindent {\bf Analytic expression of the kinetic integrals for the 
     symmetric double dot}	
 \vspace{0.5cm}

      According to Eq.(29), the kinetic integrals for the symmetric double  
      are given by the expression
         \begin{equation} 
                       K_n \!=\!   8 \gamma^2 \,  {\rm Re}  \left\{ 
                \frac{1}{A_1} {\mathcal F}_n  \left[ \frac{ [\cos (\theta/2) 
                                                \cdot (E \!-\! E_d)  +  t_d]^2}  {E-z_1}  \right] 
                  + \frac{1}{A_2}{\mathcal F}_n  [the \ same\ with \ z_2]   \right\} 
            \end{equation} 
      The first  functional in the above equation, using  Eqs.(B6),(B7),(B8), reads 
            \begin{eqnarray} 
               {\mathcal F}_n \left[ \frac{ \left[\cos (\theta/2) 
                            \, (E -E_d) +  t_d\right]^2}  {E-z_1} \right] 
               &=& \cos^2(\theta/2) \left[ k_BT\, b_{n+1} 
                   +  (\mu + z_1 - 2E_d) \, b_n \right] 
                                 \nonumber \\
                 & &  \hspace{-3.6cm} +  2\cos(\theta/2) \, t_d  b_{n} +
                  \frac{[\cos(\theta/2) \, (z_1 - E_d) + t_d]^2}{k_BT} 
                         \, J_{n}(\frac{z_1 - \mu}{k_BT})  \ . 
             \end{eqnarray} 

        It is convenient to write more explicitly the first few 
        values $F_{0,1,2}$.  Using  Eqs.(B3) we obtain the expressions:
            \begin{eqnarray*} 
               {\mathcal F}_0 \left[ \frac{ \left[\cos (\theta/2) 
                            \, (E -E_d) +  t_d\right]^2}  {E-z_1} \right] 
               &=& \cos^2(\theta/2) \, (\mu + z_1 - 2E_d) +  2\cos(\theta/2) \, t_d  
                                                    \\
                 & &  \hspace{-0.6cm}     
                 + \  \frac{ \left[\cos(\theta/2) \, (z_1 - E_d) + t_d\right]^2 }{k_BT}
                         \, J_{0}(\frac{z_1 - \mu}{k_BT})  \ .
             \end{eqnarray*}   
         Similarly: 
            \begin{eqnarray*} 
               {\mathcal F}_1 \left[ \frac{ \left[\cos (\theta/2) 
                            \, (E -E_d) +  t_d\right]^2}  {E-z_1} \right] 
               &=& \frac{\pi^2}{3}  \cos^2(\theta/2) \, k_BT   
                                            \\
                 & &  \hspace{-0.6cm}   
                 + \  \frac{  \left[\cos(\theta/2) \, (z_1 - E_d) + t_d\right]^2}{k_BT} 
                         \, J_{1}(\frac{z_1 - \mu}{k_BT})  \ .
             \end{eqnarray*}  
     It also holds
           \begin{eqnarray*} 
               {\mathcal F}_2 \left[ \frac{ \left[\cos (\theta/2) 
                            \, (E -E_d) +  t_d\right]^2}  {E-z_1} \right] 
               &=& \frac{\pi^2}{3}  \cos^2(\theta/2)  
                 \left(  \mu + z_1 - 2E_d  \right)  
                                              \\
                 & &  \hspace{-3.6cm} +  \frac{2\pi^2}{3} \cos(\theta/2) \, t_d    
                 + \frac{ \left[\cos(\theta/2) \, (z_1 - E_d) + t_d\right]^2}{k_BT} 
                         \, J_{2}(\frac{z_1 - \mu}{k_BT})  \ . \quad
             \end{eqnarray*}  
        Inserting the above result into Eq.(B9) provides the analytic expression 
        of the kinetic parameters. It holds: 
           \begin{eqnarray}
             K_0   &=&   8 \gamma^2 \,  {\rm Re} \, \left\{ \frac{1}{A_1} \left[
                \cos^2(\theta/2) \, (\mu + z_1 - 2E_d) +  2\cos(\theta/2) \, t_d  
                                        \right. \right.    \nonumber \\
                 &+&  \left. \left.     
                   \frac{ \left[\cos(\theta/2) \, (z_1 - E_d) + t_d\right]^2 }{k_BT}
                         \, J_{0}(\frac{z_1 - \mu}{k_BT})  \right] + \frac{1}{A_2}
                                     [ the \ same \ with \ z_2] \right\}   . \quad
              \end{eqnarray}  
           The next kinetic parameter reads
                 \begin{eqnarray}
             K_1   &=&   8 \gamma^2 \,  {\rm Re} \, \left\{ \frac{1}{A_1} \left[
                \frac{\pi^2}{3}  \cos^2(\theta/2) \, k_BT  
                                        \right. \right.    \nonumber \\
                 &+&  \left. \left.     
                   \frac{ \left[\cos(\theta/2) \, (z_1 - E_d) + t_d\right]^2 }{k_BT}
                         \, J_{1}(\frac{z_1 - \mu}{k_BT})  \right] + \frac{1}{A_2}
                                     [ the \ same \ with \ z_2] \right\}   . \quad
              \end{eqnarray}  
         The third kinetic parameter of interest is
             \begin{eqnarray}
             K_2   &=&   8 \gamma^2 \,  {\rm Re} \, \left\{ \frac{1}{A_1} \left[
              \frac{\pi^2}{3}  \cos^2(\theta/2) \, (\mu + z_1 - 2E_d) 
              +  \frac{2\pi^2}{3}\cos(\theta/2) \, t_d  
                                        \right. \right.    \nonumber \\
                 &+&  \left. \left.     
                   \frac{ \left[\cos(\theta/2) \, (z_1 - E_d) + t_d\right]^2 }{k_BT}
                         \, J_{2}(\frac{z_1 - \mu}{k_BT})  \right] + \frac{1}{A_2}
                                     [ the \ same \ with \ z_2] \right\}   . \quad
              \end{eqnarray}  
           For convenience the  basic results for   actual simulations 
           are summarized in Table I.


 \vspace {1.5cm}
 {\bf Acknowledgments}
The authors acknowledge the ``IT center" of the University of Pisa for the computational support.

  \vspace{1cm}

  \end{document}